\documentclass{elsart}

\usepackage{rotating}
\usepackage{epsfig}
\usepackage{cite}

\usepackage{amssymb}

\def\mum{$\mu$m}
\def\be{\begin{equation} }
\def\ee{\end{equation} }
\def\ba{\begin{eqnarray} }
\def\ea{\end{eqnarray} }
\def\ban{\begin{eqnarray*} }
\def\ean{\end{eqnarray*} }

\begin{document}

\begin{frontmatter}


\title{Position Sensing from Charge Dispersion in 
Micro-Pattern Gas Detectors with a Resistive Anode}

\author[ref_CU,ref_TR]{M.~S.~Dixit\corauthref{add_MSD}},
\ead{msd@physics.carleton.ca}
\author[ref_JD]{J.~Dubeau},
\author[ref_UM]{J.-P.~Martin}
and
\author[ref_CU]{K.~Sachs}

\address[ref_CU]{Department of Physics, Carleton University, 
        \\ 1125 Colonel By Drive, Ottawa, ON K1S 5B6 Canada}
\address[ref_JD]{DETEC, Aylmer, QC, Canada}
\address[ref_UM]{Universiy of Montreal, Montreal, QC, Canada}
\address[ref_TR]{TRIUMF, Vancouver, BC Canada}
\corauth[add_MSD]{Corresponding author.
         Tel.: +1-613-520-2600, ext. 7535; fax: +1-613-520-7546.}


\begin{abstract}
 Micro-pattern gas detectors, such as the Gas Electron Multiplier (GEM) and the Micromegas 
need narrow high density anode readout elements to achieve good spatial resolution. A 
high-density anode readout would require an unmanageable number of electronics channels for certain 
potential micro-detector applications such as the Time Projection Chamber. We describe below a 
new technique to achieve good spatial resolution without increasing the electronics channel 
count in a modified micro-detector outfitted with a high surface resistivity anode readout 
structure. The concept and preliminary measurements of spatial resolution from charge 
dispersion in a modified GEM detector with a resistive anode are described below.
\end{abstract}

\begin{keyword}
Gaseous Detectors \sep 
Position-Sensitive Detectors \sep
Micro-Pattern Gas Detectors \sep
Gas Electron Multiplier \sep
Micromegas

\PACS 29.40.Cs \sep 29.40.Gx 

\end{keyword}
\end{frontmatter}

\section{Introduction}
\label{sec:intro}

  A new class of high-resolution multi-channel gas avalanche micro-detectors has been 
developed during the past decade for charged particle tracking. The Gas Electron Multiplier 
(GEM) \cite{cit:GEM} and the Micromegas \cite{cit:MM} are examples of some of the new micro-pattern gas 
detectors \cite{cit:mpgd} already in wide use. The GEM and the Micromegas sample the avalanche charge 
using arrays of closely spaced long parallel anode strips to measure a single co-ordinate.  
Spatial resolutions of 40 to 50 \mum\ are typical with anodes strips at 200 \mum\ pitch. 
Micro-detectors do require more instrumented channels of electronics than multi-wire proportional 
chambers. However, the number of readout channels has not yet become an issue for most 
experiments using micro-detectors for charged particle tracking. 

 There are potential micro-detector applications, however, where the electronics channel count 
may become an issue. The Time Projection Chamber (TPC) \cite{cit:TPC} used in high energy physics 
experiments is one such example. A single endcap detector is used in the TPC to measure 
both the radial and the azimuthal co-ordinates of the ionization charge cluster. In the 
conventional TPC, read out with a multi-wire proportional chamber endcap, the radial 
co-ordinate is obtained from the anode wire position. A precise second co-ordinate along the 
anode wire length is measured by sampling the induced cathode charge with a series of 
several mm wide rectangular pads.

 The spatial resolution of a TPC in a high magnetic field is dominated by the 
{\boldmath $E\times B$\unboldmath} and track 
angle systematic effects \cite{cit:clif,cit:amend}. Replacement of the usual anode wire/cathode pad readout with 
one based on micro-detectors with anode pad readout would almost entirely eliminate the 
systematics and has the potential to improve the TPC performance significantly. However, the 
suppression of the transverse diffusion in a high magnetic field may often result in the 
collection of most of the avalanche charge within the width of a single anode pad resulting in 
a loss of TPC resolution. For better resolution, a micro-detector readout TPC will need either 
a finely segmented anode pad structure with a prohibitively large number of instrumented 
channels of electronics, or perhaps the complication of specially shaped pads which enhance 
anode charge sharing \cite{cit:chevron}. 

 We describe here a new technique which can be used to measure the position of a localized 
charge cluster in a micro-detector using pads of widths similar to those used in wire-pad 
systems. Most of our tests have so far been done with a modified double-GEM detector. 
However, the new technique appears to be sufficiently general to be applicable to the 
Micromegas.

\section{Charge dispersion in a micro-detector with a resistive anode}

\begin{figure}[bp]
\centerline{\mbox{\epsfxsize=\textwidth \epsffile{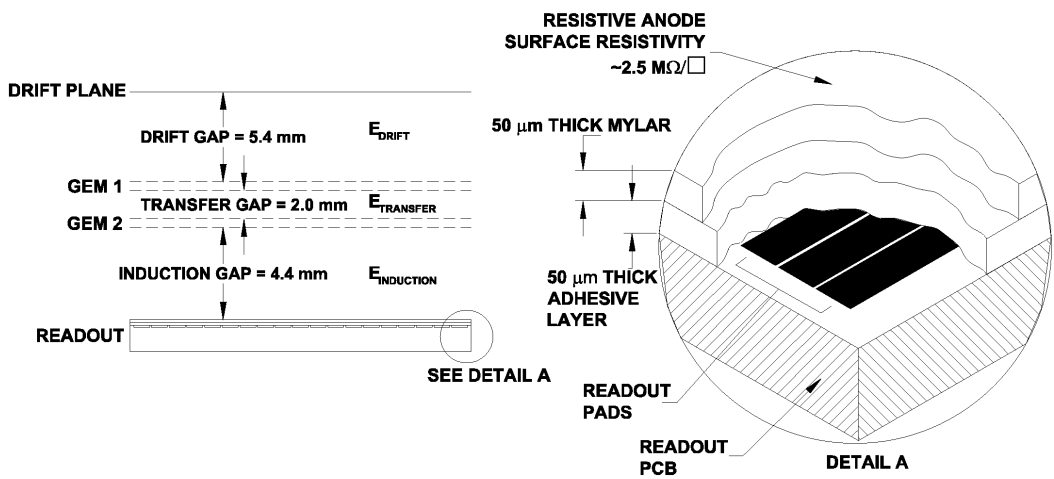}}}
\caption[]{\label{fig:scheme}
Schematics of the resistive anode double-GEM detector used for charge dispersion 
studies.
}
\end{figure}

 With certain modifications to the anode readout structure, it is possible to measure the position 
of a localized charge cluster in a micro-detector with pads wider than have been used so far. A 
thin high surface resistivity film is glued to a separate readout pad plane and is used for the 
anode (Fig. \ref{fig:scheme}). The resistive anode film forms a distributed 2-dimensional resistive-capacitive 
network with respect to the readout pad plane. Any localized charge arriving at the anode surface 
will be dispersed with the $RC$ time constant determined by the anode surface resistivity and the 
capacitance density determined by the spacing between the anode and readout planes. With the 
initial charge dispersed and covering a larger area with time, wider pads can be used for signal 
pickup and position determination. 

This spatial dispersion, which can be explained by a simple physical
model described below, does not attenuate the pad charge signal.
As the initial charge cluster diffuses, the electron density on the 
resistive surface, varying with time, is capacitively mirrored on the readout 
pads below. Since the readout pads are directly connected to the preamplifiers, 
there is no charge signal amplitude loss.

\subsection{A model for charge dispersion in a micro-detector with a resistive anode}

 The resistive anode and the readout plane together can be looked upon to form a distributed 
2 dimensional $RC$ network in the finite element approximation. Consider first the 1 dimensional 
problem of a point charge arriving at $t=0$ at the origin in the middle of an infinitely long wire 
grounded at both ends.  For small inductance, the space-time evolution of the charge density 
$\rho$ on the wire is given by the well-known Telegraph equation:
\be
\frac{\partial \rho}{\partial t} = h \frac{\partial^2 \rho}{\partial x^2}
\hspace{1cm}\mbox{where} \hspace{1cm}
h = 1/RC \; . \label{eq:T-1dim} \ee
Here $R$ is resistance per unit length and $C$ the capacitance per unit length for the wire. 

The solution for charge density is given by:
\be
\rho(x,t) = \sqrt{\frac{1}{4\pi t h}} \exp(-x^2 / 4 t h)\; . \label{eq:1dim}  \ee

 In analogy with the 1 dimensional case, we can write the Telegraph equation for the case of a 
resistive surface. At time $t = 0$, a point charge is collected at the origin by a resistive anode 
surface of infinite radius (for simplicity). The 2-dimensional Telegraph equation for the charge 
density is:
\be
\frac{\partial \rho}{\partial t} = h \left[ \frac{\partial^2\rho}{\partial r^2} +
\frac{1}{r}\frac{\partial \rho}{\partial r} \right] \; , \label{eq:T-2dim}  \ee
where in this case, $R$ is the surface resistivity and $C$ is capacitance per unit area.

The solution for the charge density function in this case is given by:
\be
\rho(r,t) = \frac{1}{2 t h} \exp(-r^2 / 4 t h)\; . \label{eq:2dim}  \ee

\begin{figure}[bp]
\centerline{\mbox{\epsfxsize=\textwidth \epsffile{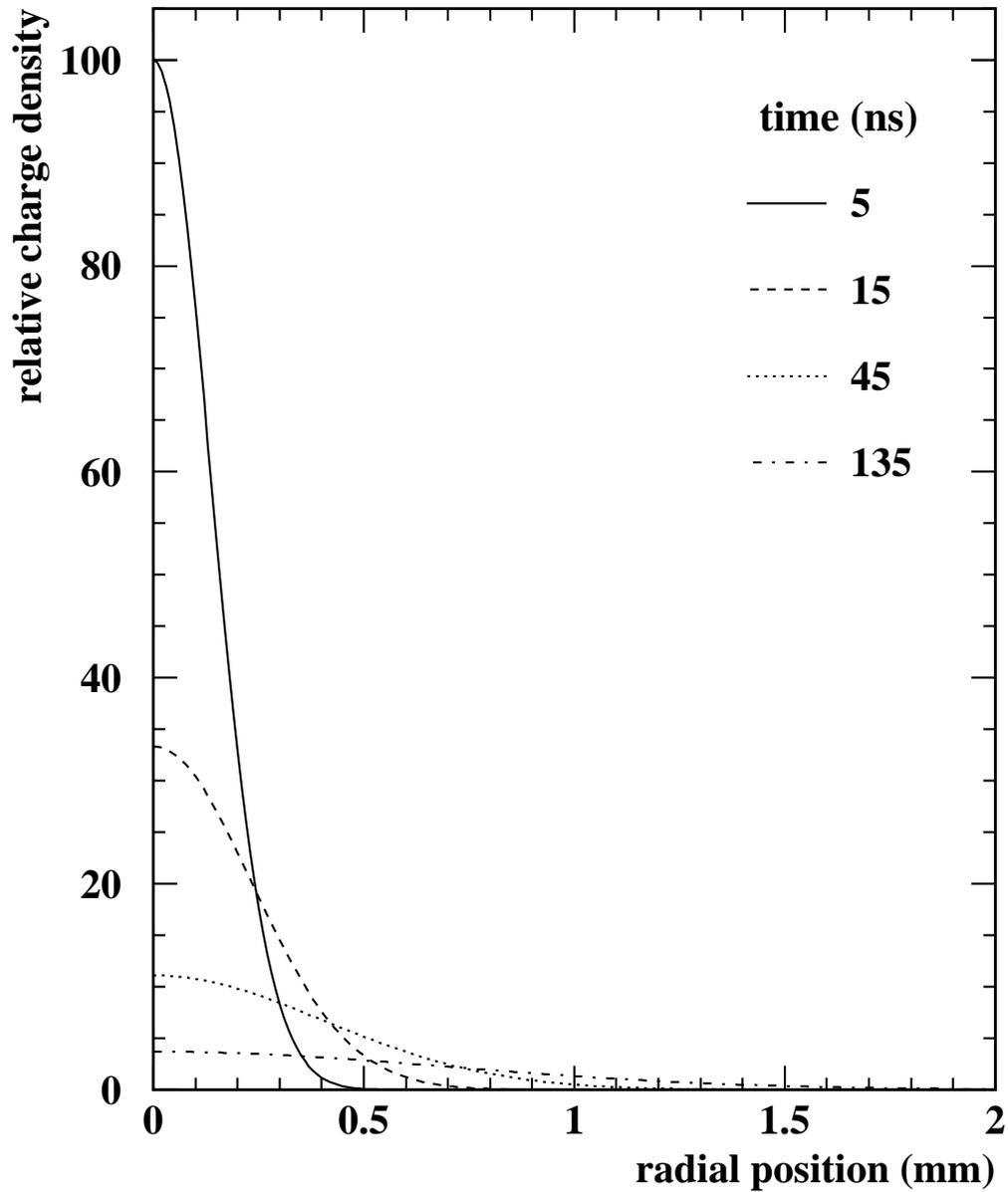}}}
\caption[]{\label{fig:charge}
The evolution of the two dimensional charge density function (equation \ref{eq:2dim}) 
on the resistive anode with a resistivity of 2.5~M$\Omega$ per 
square and anode-readout plane separation of 100~\mum . 
For the model calculation the initial 
charge was point-like and localized at the origin at time = 0.
}
\end{figure}

The charge density function (equation \ref{eq:2dim}) for the resistive anode varies with time and is 
capacitively sampled by the readout pads. Fig. \ref{fig:charge} shows the time evolution of the charge density 
for an initially localized charge cluster for our detector. The charge 
signal on a pad can be computed by integrating the time dependent charge density function over 
the pad area. The shape of the charge pulse on a pad depends on the pad geometry, the location 
of the pad with respect to the initial charge and the $RC$ time constant of the system.

\subsection{Charge dispersion signal in micro-detectors with long readout strips}
 
The charge dispersion measurements were carried out with a modified GEM detector with long 
anode strips. Since a spatial co-ordinate measurement for long strips is meaningful only in a 
direction transverse to the strip length, we can use the 1-dimensional Telegraph equation 
(\ref{eq:T-1dim}) to 
describe the situation. However, the solution for the charge density must account for the finite 
size of the resistive anode in contrast to the solution given by equation (\ref{eq:1dim}) in the long wire 
approximation. 

The boundary conditions to solve equation (\ref{eq:T-1dim}) in this case are: 
\be
\rho(x=0,t) = \rho(x=s,t) = 0\; ; \; 0 \leq t \leq \infty  \; , \label{eq:boundary}  \ee
where $s$ is the size of the resistive foil (assumed square) held at ground potential along its 
boundary.

The solution satisfying the finite boundary conditions is:
\be
\rho(x,t) = \sum_{m=1}^\infty A_m \exp [ -h(m\pi/ s)^2 t] \sin(xm\pi/s) \; , \label{eq:simsol}  \ee
where the coefficients $A_m$ can be determined from the knowledge of the initial charge density:
\be
A_m = \frac{2}{s}\int_0^s \rho(x,t=0) \sin(xm\pi/s) dx \; . \label{eq:coeff}  \ee

 The signal on a readout strip can be computed by integrating the charge density function over 
the strip width. Furthermore, the finite extent of the initial charge cluster, the intrinsic 
micro-detector charge signal rise time as well as the rise and fall time characteristics of the front-end 
electronics determine the shape of the measured signal shape. All these parameters need to be 
included in the model to compare to the experiment.
 
\begin{figure}[bp]
\centerline{\mbox{\epsfxsize=\textwidth \epsffile{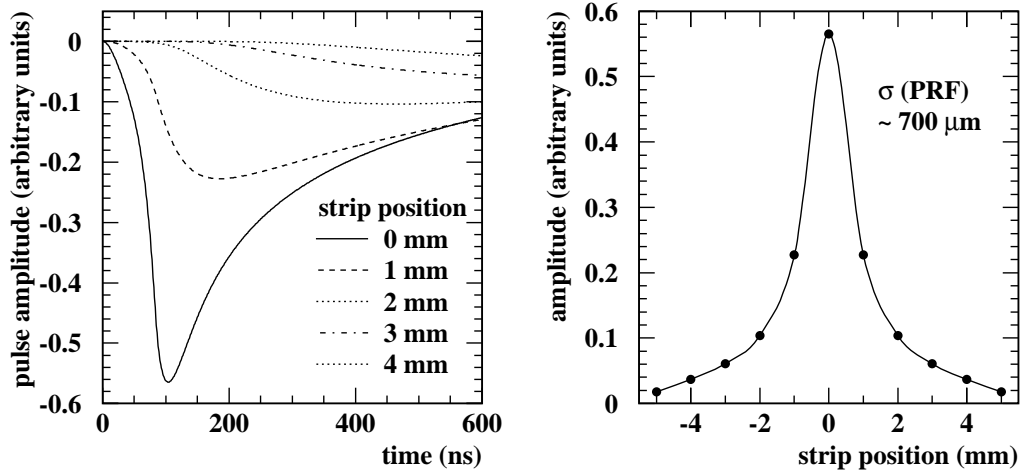}}}
\caption[]{\label{fig:simsig}
Simulated signals for 1 mm wide strips for an anode resistivity of
2.5~M$\Omega$ per 
square and anode-readout plane separation of 100~\mum . The pad response function for charge 
dispersion is shown on the right.  The diffusion effects were neglected for this simulation.
}
\end{figure}

Model calculations were done for a GEM detector with a resistive anode readout with 1 mm 
wide strips. The anode resistivity and anode-readout gap limits
the computed spatial spread of the charge dispersion over pads to about 700~\mum\ comparable to 
transverse diffusion in a high magnetic field TPC. Simulated signals for the readout strip directly 
below the initial ionization charge cluster and for the next four adjacent strips are shown in 
Fig. \ref{fig:simsig}. The same figure also shows the simulated pad response function or equivalently the 
spatial spread of an initially localized charge cluster due to charge dispersion.

\section{Spatial resolution measurements in a GEM detector with a resistive anode}
 
The charge dispersion test measurements were made with the modified double-GEM detector 
(Fig. \ref{fig:scheme}) filled with Ar/CO$_2$ 90/10. A 50 \mum\ thick Mylar film with a 
surface resistivity of 2.5 M$\Omega$
per square was glued to the pad readout board. The spacing between the anode and readout 
planes was close to 100~\mum\ including the thickness of the glue. 

The initial ionization is provided by x-ray photon conversion in the gas. 
The average photon energy was 
about 4.5 keV as low energy bremsstrahlung photons from the copper target x-ray tube, run at 7 
kV, were absorbed by the material in the x-ray tube and detector windows. A 
\mbox{$\sim$40 \mum} pinhole in 
a thin brass sheet was used to produce a miniaturized x-ray tube focal spot image in the GEM 
drift gap. The size of the x-ray spot at the detector is estimated to be on the order of 
70 \mum . 
After avalanche multiplication and diffusion, the RMS size of the electron cloud reaching the 
resistive anode was \mbox{$\sim$400 \mum}. The gas gain was about 2500.

Signals were read out from 7 cm long and 1.5 mm wide strips. The front-end electronics 
consisted of Aleph \cite{cit:amend} TPC wire charge preamplifiers followed by receiver amplifiers. Signals 
from 8 contiguous strips were digitized using two 4-channel Tektronix digitizing oscilloscopes. 
A computerized translation stage was used to move the x-ray spot in small steps over the width 
of the centre strip. One thousand event runs were recorded for each x-ray spot position on an 
event by event basis. 

\begin{figure}[bp]
\epsfxsize=\textwidth \epsffile{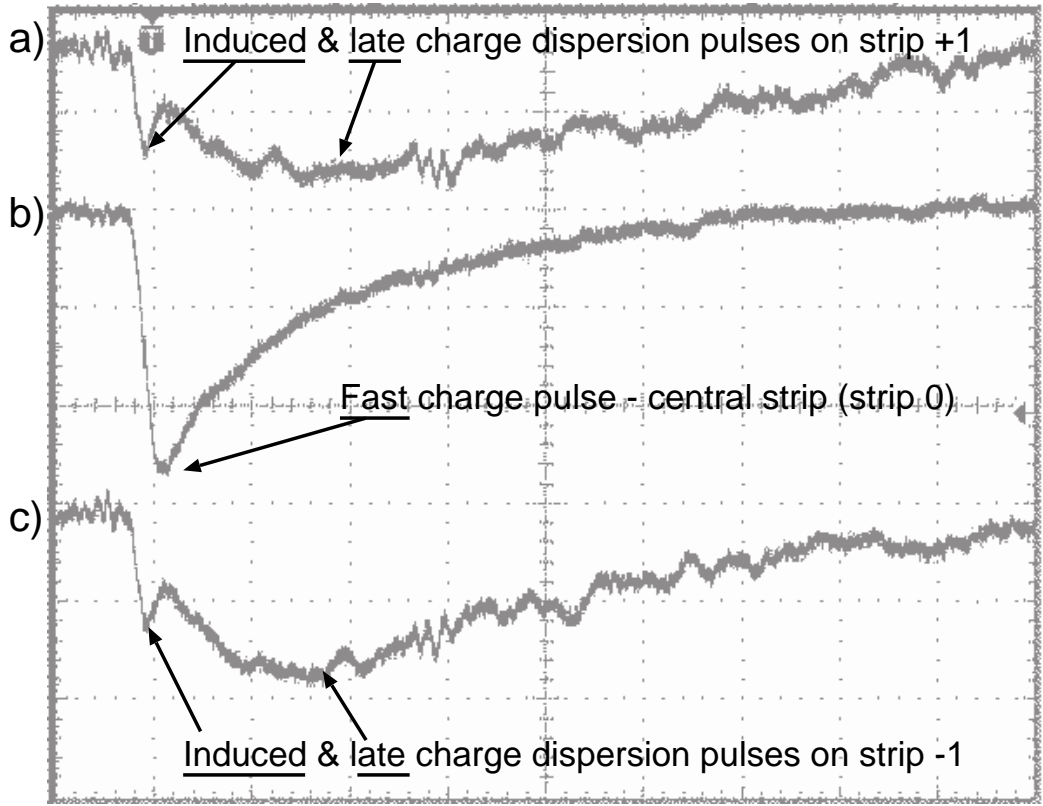}
\caption[]{\label{fig:scope}
Observed charge dispersion signals on three adjacent strips for a single x-ray photon conversion in the 
double-GEM detector. Tektronix scope pulses with \mbox{400 ns/div} on the 
a)~right strip (20 mV/div), 
b)~central strip (50 mV/div) and
c)~left strip (20 mV/div).}
\end{figure}

For a given anode surface resistivity and readout geometry, the observed shape of the charge 
pulse depends on the strip position with respect to the location of primary charge cluster on the 
resistive anode.  Fig. \ref{fig:scope} shows an event where the x-ray ionization spot is located directly above 
the centre of a readout strip. A fast charge pulse is observed on the strip peaking in time with the 
maximum of the charge density at the anode surface above. Pulses on strips farther away have a 
slower rise time and peak late because the local charge density on the anode surface nearest the 
strip reaches its maximum later. Also, an early short duration induced pulse is visible for strips 
adjacent to the main strip. The induced pulses \cite{cit:orsay}, produced by electron motion in the GEM 
induction gap, have demonstrated position sensitivity \cite{cit:LCWS} but require the use of high-speed pulse 
shape sampling electronics. In addition, measurable induced pulses are specific to GEM 
detectors with sizeable induction gaps. For charge dispersion measurements described below, the 
induced GEM pulse information has not been used.

\section{Data analysis and results}

The charge dispersion signals were confined to a narrow region on the readout board in the 
present setup. There were measurable signals above noise only on three 1.5 mm wide strips. The 
analysis to determine the space point resolution from the event by event data consisted of 
following steps: a) Determine the pulse heights of signals on the strips, independent of rise time; 
b) Compute a centre of gravity position for the event from the measured pulse heights; c) Correct 
for the bias in the computed centre of gravity to obtain the position of the ionization cluster for 
the event. The standard deviation of the bias corrected centre of gravity position with respect to the 
known x-ray beam spot position for the run gives a measure of the space point resolution. 

\begin{figure}[bp]
\centerline{\mbox{\epsfxsize=\textwidth \epsffile{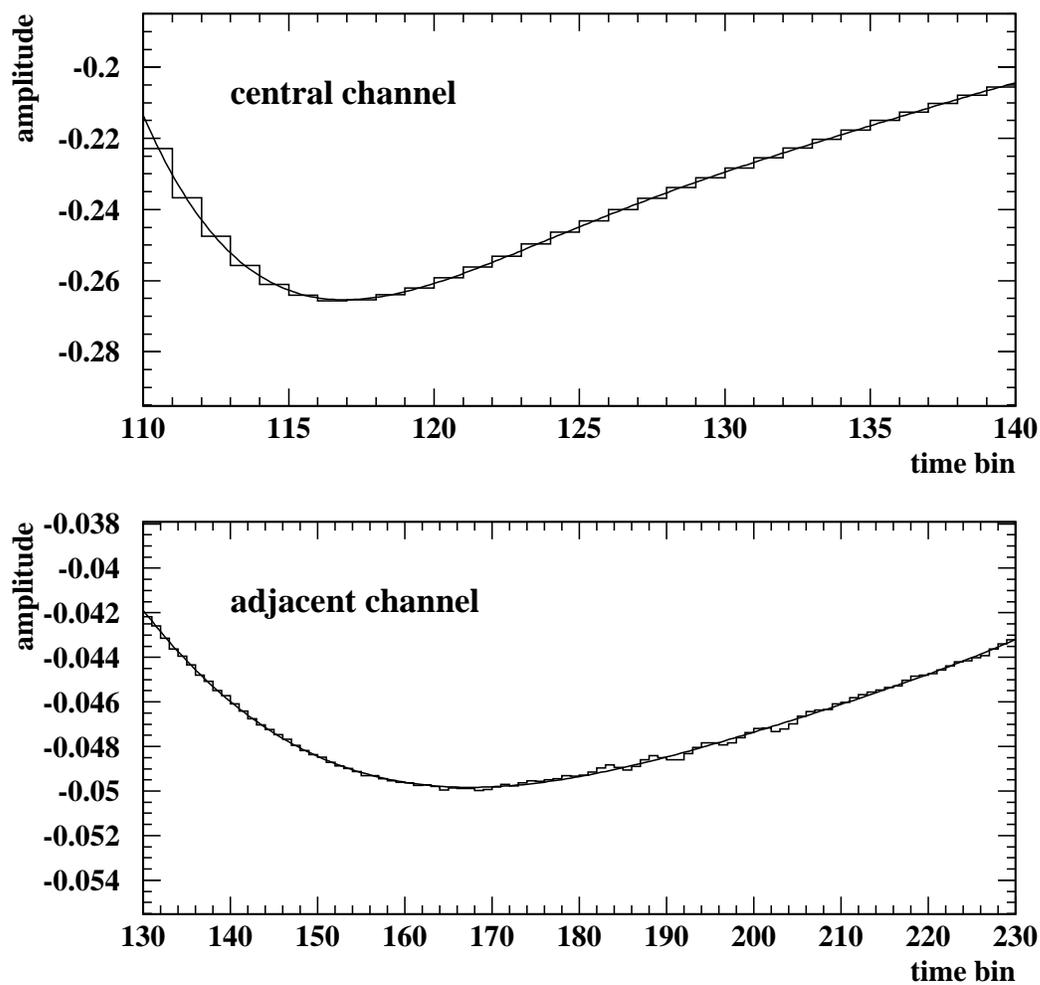}}}
\caption[]{\label{fig:poly}
Polynomials were fitted to the calibration data set for each run to fix the functions used to 
determine the pulse height of events for resolution studies. The top figure shows the polynomial 
fit to the average of fast charge pulses for the centre strip. The figure below shows the 
polynomial fit to the average of late charge dispersion pulses for an adjacent strip.
}
\end{figure}

The data from each 1000 event run were sub-divided into two equal data sets: one used for 
calibration and one for resolution studies. The pulse heights were obtained by fitting polynomial 
functions to the digitized pulse shape data. The calibration data set was used to determine and fix 
the coefficients of polynomial functions (see Fig. \ref{fig:poly}) used subsequently in the analysis of events 
in the resolution data set. 

\begin{figure}[bp]
\centerline{\mbox{\epsfxsize=10cm \epsffile{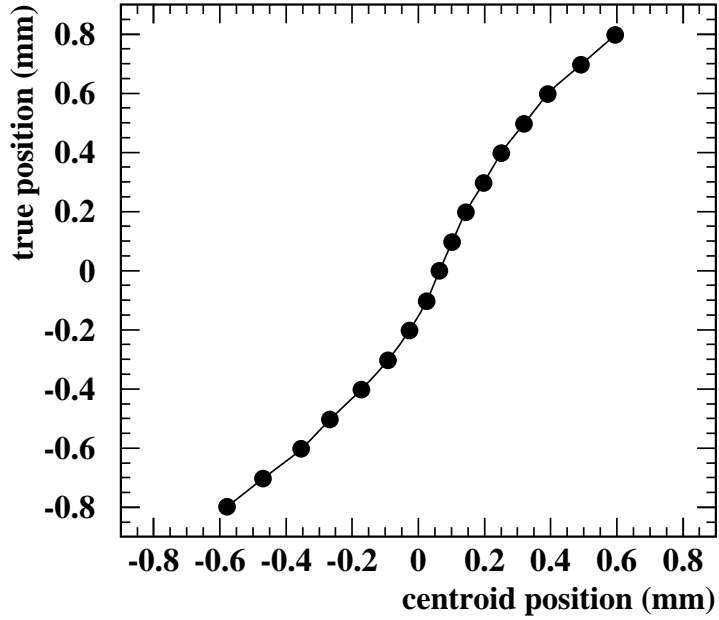}}}
\caption[]{\label{fig:bias}
The bias correction function shown in the figure was experimentally determined. The bias 
function was used in converting the computed centroid of signals on three strips to the true 
position on an event by event basis. 
}
\end{figure}

The centres of gravity for the events in the calibration data set were computed from the measured 
pulse heights. The correction function for the bias in the centre of gravity method 
(see Fig. \ref{fig:bias}) 
was determined by plotting the mean value of computed centres of gravity against the known 
x-ray spot position for the individual runs. 

\begin{figure}[bp]
\centerline{\mbox{\epsfxsize=\textwidth \epsffile{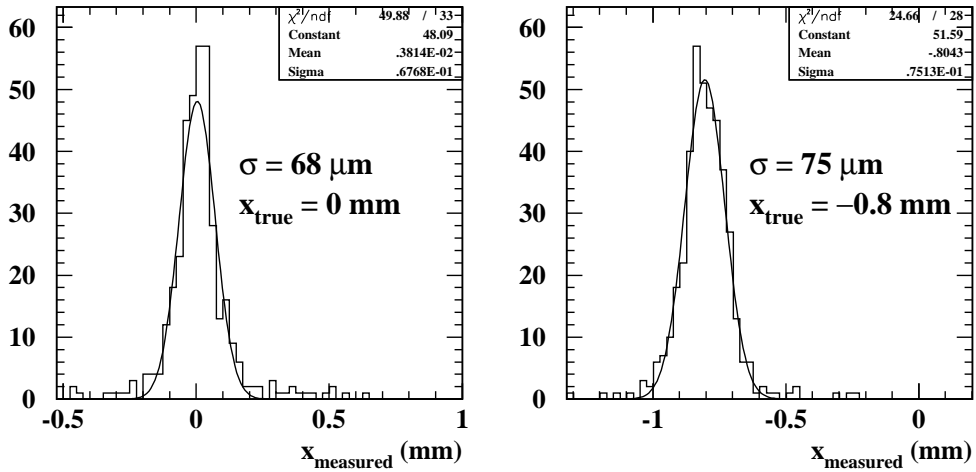}}}
\caption[]{\label{fig:resolution}
The resolution function for two x-ray beam spot positions.
}
\end{figure}

\begin{figure}[bp]
\centerline{\mbox{\epsfxsize=\textwidth \epsffile{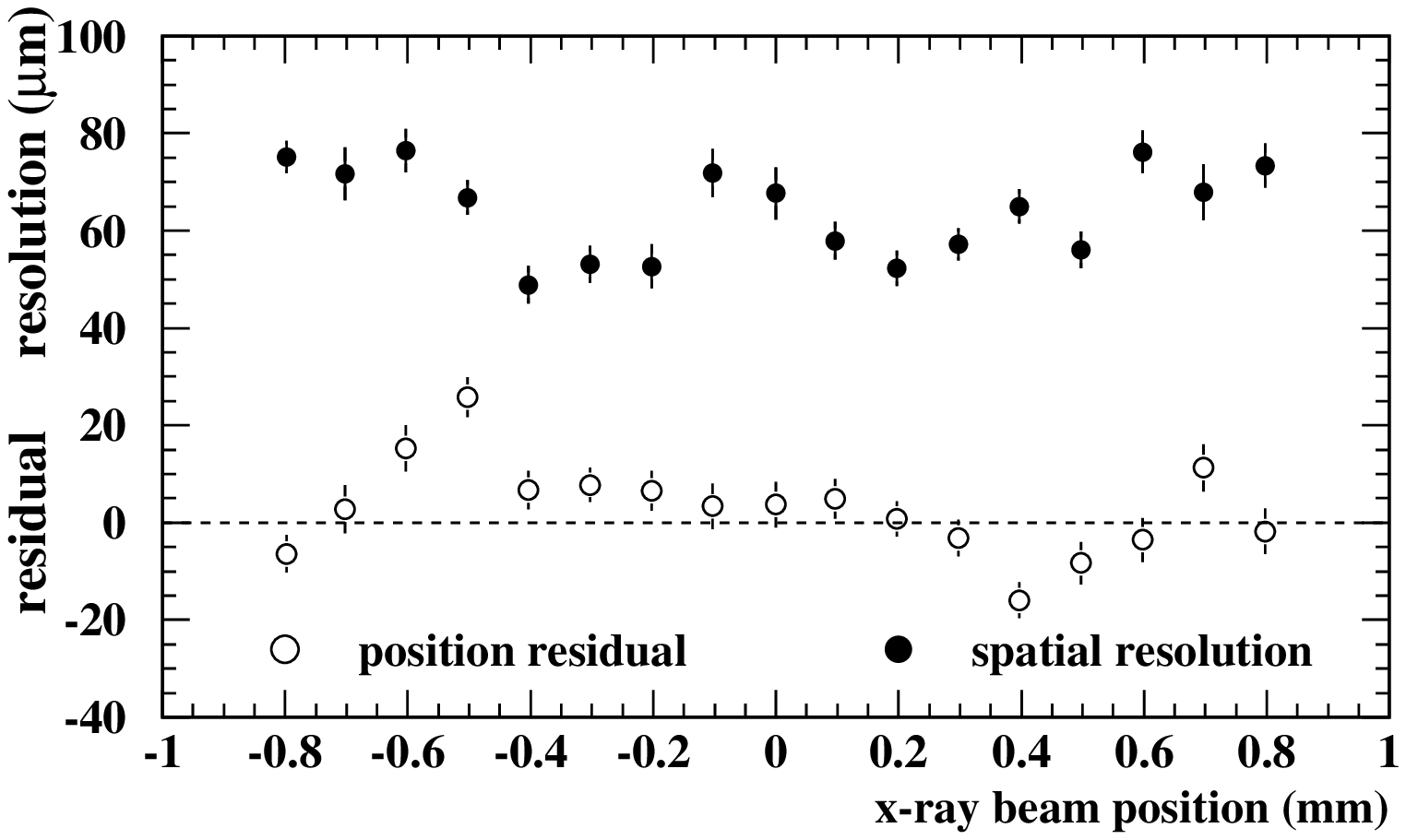}}}
\caption[]{\label{fig:summary}
The summary of the spatial resolutions and the position residuals for the x-ray scan across 
the readout strips.
}
\end{figure}

The pulse heights and the peak positions for the events used for resolution studies were 
determined by fitting the pulses to the fixed polynomials shapes obtained from the corresponding 
calibration data set. The computed centres of gravity were converted to ``true position'' by 
interpolation using the bias correction function determined from the calibration data set. 
Fig. \ref{fig:resolution} 
shows the measured resolution function for the 1.5 mm wide strips at two different positions of 
the x-ray ionization spot over the strip. Fig. \ref{fig:summary} shows the measured spatial resolutions and the 
position residuals: i.e. the deviation of measured positions from the micrometer settings. The 
small systematic trends apparent in the plot appear to be related to an imprecise knowledge of the 
experimental parameters in the present setup and any remaining biases in the present method of 
analysis. Nevertheless, the standard deviations of the position measurements, in the range of 50 
to 80 \mum , are all consistent with the size of the collimated x-ray spot at the detector.

\section{Outlook and summary}

We have demonstrated that a controlled $RC$ dispersion of the avalanche charge makes it possible 
to measure its position with a micro-detector with strips wider than have been used previously. 
The pad response function and signal shapes are determined by the anode surface resistivity and 
anode-readout plane gap. With the proper choice of the $RC$ time constant of the system, the 
charge dispersion technique will not compromise the counting rate ability of the detector. Nor 
should it compromise the 2-track resolving power of the detector which should be limited only 
by the diffusion effects in the gas. Once the characteristics of the charge dispersion signal are 
properly understood, it should also be possible to simplify the technique replacing the pulse 
shape measurement system with less expensive charge integrating electronics. 

 The charge dispersion space point resolution studies described here were done for long readout 
strips in a modified GEM detector with a resistive anode. Further experimental and simulation 
studies are in progress to investigate the spatial resolution capabilities of the charge dispersion 
technique with rectangular pads similar to those used in wire-cathode pad TPCs. We are also 
testing the concept of position sensing from charge dispersion with the Miromegas where the 
high anode resistance may help improve the detector HV stability as well as protect the front-end 
electronics from spark damage.

\clearpage

\section*{Acknowledgments}
 
The charge preamplifiers used in these measurements came from the Aleph TPC at CERN and 
we wish to thank Ron Settles for making these available to us. Ernie Neuheimer lent us his 
expertise in designing, building and troubleshooting much of the specialized electronics used for 
these measurements. Mechanical engineers Morley O'Neill and Vance Strickland helped with 
the detector design and in improving the clean-room facility used for the detector assembly. 
Philippe Gravelle provided technical assistance when needed. Our CO-OP students Alasdair 
Rankin and Steven Kennedy made significant contributions to all aspects of this research from 
hardware construction to improving the data acquisition software as well as writing some of the 
data analysis code. Finally, one of the authors (MSD) would like to express his thanks to 
V.~Radeka for an illuminating discussion concerning the phenomenon of charge dispersion. This 
research was supported by a project grant from the Natural Sciences and Engineering Research 
Council of Canada.

\clearpage

\begin{thebibliography}{00}
\bibitem{cit:GEM} F. Sauli, Nucl. Inst. Meth. {\bf A386} (1997) 531.
\bibitem{cit:MM} Y. Giomataris et al, Nucl. Inst. Meth. {\bf A376} (1996) 29.
\bibitem{cit:mpgd} F. Sauli and A. Sharma, Ann. Rev. Nucl. Part. Sci. {\bf 49} (1999) 341.
\bibitem{cit:TPC} D. R. Nygren, PEP 198 (1975).
\bibitem{cit:clif} C.K.Hargrove et al, Nucl. Inst. Meth. {\bf 219} (1984) 461.
\bibitem{cit:amend} S. R. Amendolia et al., Nucl. Inst. Meth. {\bf A283} (1989) 573. 
\bibitem{cit:chevron} M. Schumacher, LC-DET-2001-014.
\bibitem{cit:orsay} M.S. Dixit et al, Proceedings of Workshop on Micro-Pattern Gas Detectors, Orsay France 
(1999).
\bibitem{cit:LCWS} D. Karlen et al, Physics and experiments with future linear $\rm e^+e^-$ colliders, LCWS2000, 
American Institute of Physics Conf. Proc. Vol 578.

\end{thebibliography}
\end{document}